\documentclass[12pt]{article}
\begin{document}
\begin{flushright}
CTP-TAMU-42/97\\
ACT-15/97
\end{flushright}
\vspace{12pt}
\thispagestyle{empty}
\begin{center}
{\Large\bf M-Phenomenology}\footnote{Based on invited talks delivered 
during the summer, 1997.}\\
\vspace{36pt}
{D. V. Nanopoulos}\\
\vspace{24pt}
{\small\it Center for Theoretical Physics, Department of Physics\\
Texas A\&M University, College Station, TX 77843--4242, USA\\
Astroparticle Physics Group, Houston Advanced Research Center\\ 
The Mitchell Campus, The Woodlands, TX 77381, USA\\
Academy of Athens, Chair of Theoretical Physics, Division of Natural Sciences\\
28 Panepistimiou Avenue, Athens 10679, Greece }
\end{center}
\begin{abstract}
Recent developments involving strongly coupled superstrings are discussed 
from a phenomenological point of view.
In particular, strongly coupled $E_8\times E'_8$ is described as an 
appropriate long-wavelength limit of M-theory, and some generic 
phenomenological implications are analyzed, including a long sought
downward shift of the string unification scale and a novel way to 
break supersymmetry. A specific scenario is presented that leads to 
a rather light, 
and thus presently experimentally testable, sparticle spectrum.
\end{abstract}
\section{Introduction}
The standard model (SM) of particle physics, an 
$SU(3)_c\times SU(2)_L\times U(1)_Y$ gauge group with the appropriate 
field representations, seem to fit all presently available
experimental data, including, notably, the LEP high precision
electroweak tests\cite{Altarelli:1996eq}.
Most remarkably, a supersymmetric extension of the 
standard model (SSM), while it more than doubles the SM particle
content in the mass range 
(${\cal O}(100{\rm GeV}\rightarrow 1{\rm TeV})$), 
it does not only escape unscathed from all the LEP
severe tests\cite{Altarelli:1996eq}, 
but it provides the first evidence for
Superunification\cite{Ellis:1990zq}.
Indeed, renormalization group extrapolation of the
three coupling constants ($\alpha_3$, $\alpha_2$, $\alpha_1$), as
measured at LEP, at very high energy show that they do converge at
some scale  
\begin{equation}
M_{\rm GUT}\equiv M_{\rm LEP}\simeq
{\cal O}(10^{16}{\rm GeV}),
\label{eq:mgut}
\end{equation}
at a common value
\begin{equation}
\alpha_{{\rm GUT}}\simeq \frac{1}{25},
\label{eq:alphagut}
\end{equation}
as theoretically predicted a long time ago\cite{Dimopoulos:1991au}.
While very suggestive, this value of the Grand Unified scale brings us
suspiciously close to the SPlanck scale:
\begin{equation}
M_{\rm SPl}\equiv \frac{M_{\rm Pl}}{\sqrt{8\pi}}=
\frac{1}{\sqrt{8\pi G_N}}\simeq 2.4\times 10^{18}{\rm GeV},
\label{eq:splanck}
\end{equation}
implying that gravitational effects may be non-negligible and should
be taken into account.

Supergravity is the local extension of rigid supersymmetry (SUSY),
that automatically involves gravity. While Supergravity (SUGRA) cannot
provide a finite quantum field theory, and thus  a consistent 
quantum theory of
gravity, still it may serve as an {\it effective theory} for energy 
scales below the SPlanck scale (\ref{eq:splanck}). 
Usually, SUGRA models are
plagued by a major catastrophe, namely by  an unacceptably large value
for the cosmological constant, $\Lambda_c$,  at the
classical level. Interestingly enough, there is a specific class of
models consisting of the so called 
{\it no-scale supergravity framework}\cite{Lahanas:1987uc}, that 
\begin{itemize}
\item provides a naturally vanishing cosmological constant,
$\Lambda_c$, {\it at least} at the classical level due to the available
flat directions ($T_i$) of the scalar potential: 
$V(T_i) = 0$\cite{Cremmer:1983bf}
\item supergravity is spontaneously broken, but the gravitino mass,
$m_{3/2}$, is {\it undetermined} at the classical level:\cite{Cremmer:1983bf}
\begin{equation}
m_{3/2}=m_{3/2}(T_i)\ne 0.
\label{eq:m32}
\end{equation}
\item quantum corrections curve, in principle, the flat directions of the
scalar potential $V$, thus creating {\it dynamically} a $V_{{\rm min}}$,
and provide vev's to the $T_i$'s. In other words, we have not only
succeeded to get radiative electroweak breaking (REWB), but we do also
have a dynamical determination of the SUGRA breaking scale: 
$m_{3/2}=m_{3/2}(\langle T_i\rangle)$. 
In principle, in the no-scale
supergravity, as it is suggested by its name, all mass scales are
dynamically determined in terms of a single scale (our yardstick), say
the SPlanck scale\cite{Ellis:1984sf,Ellis:1984ei,Ellis:1984bm}.
\item the flat directions of the scalar potential $V$, corresponding to
usually called moduli fields $T_i$ trace their origin to the existence of
non-compact continuous global symmetry, duality group, {\it e.g.}
$SU(1,1)$, abundant in extended supergravity 
theories\cite{Cremmer:1983bf,Ellis:1984ei,Ellis:1984bm}.
\end{itemize}
These rather unique characteristics find their natural habitat in
string theory, whose infrared limit is nothing else but no-scale
supergravity\cite{Witten:1985xb}. Sections 2 and 3 describe the weak
and strong coupling limits of superstrings, respectively, while 
section 4 provides a phenomenological profile of strongly coupled 
$E_8\times E'_8$ viewed as an appropriate limit of M-theory. M-theory 
inspired supersymmetry phenomenology is discussed in section 5 and 
some conclusions are drawn in section 6

\section{Superstrings: Weak Coupling (10-D $\rightarrow$ 4-D)}
Superstrings, one-dimensional extended objects, provide a consistent
quantum theory of gravity and a natural framework for realistic
unification of all fundamental interactions. While superstrings are
intrinsically different from point-like particles in many respects,
one may still, at least, initially try to use perturbation theory in
order to get out some physics. The initial stages of this
(perturbative) programme has yielded rather interesting results 
\begin{itemize}
\item useful gauge groups are available, $SU(3)_c\times SU(2)_L\times
U(1)_Y$, $SU(5)\times U(1)$, \ldots
\item useful particle content is present, fitting into highly desirable
representations of the above gauge groups, such as, three generations
of quarks and leptons, two Higgs doublets, etc.
\item useful superpotential form, yielding among other things
successful Yukawa couplings, {\it e.g.} 
\begin{equation}
\lambda_t\simeq g^2\sim 0.7,
\label{eq:0.7}
\end{equation}
at the string scale implying a top quark 
mass\cite{Lopez:1994iu,Lopez:1996zz}
\begin{equation}
m_t\simeq 160-190{\rm GeV},
\label{eq:mt} 
\end{equation}
or successful fermion mass relations\cite{Lopez:1991ac}, {\it e.g.},
\begin{equation}
\frac{m_c}{m_t}\simeq \frac{1}{2}(\frac{m_e}{m_\tau})^{1/2}.
\label{eq:mrelation}
\end{equation}
More generally, the ``technology'' has been developed to understand
the so called, in the late 70's, generation problem and the fermion
mass problem at the most fundamental level of dynamics, {\it i.e.} at
the string scale. Dynamically calculable 
``{\it non-renormalizable}'' terms\cite{Kalara:1991sq}, may
provide the rather intrigued textures needed to explain the observed
fermion mass spectrum.
\item Supersymmetry emerges {\it naturally} 
due to the highly constrained form
of string dynamics, in other words supersymmetry is a prediction of
string theory. Furthermore, the low-energy limit of string theory is,
generically, described by an effective theory belonging to the
no-scale supergravity framework\cite{Lahanas:1987uc}.
\end{itemize}

Despite all the above remarkable, indeed, successes of weakly coupled
strings, there are enough stumbling blocks in the way towards a
realistic superunification, that shed shadows of doubt on the whole
picture, such as
\begin{itemize}
\item it has not delivered yet a unique
theory at the string scale. Actually, the problem is
not that we are short of ``theories''. On the contrary, we seem to
have more than it is necessary. Instead of {\it one}, we got  
{\it five} theories: type I, type IIA, type IIB, heterotic SO(32), and
heterotic $E_8\times E'_8$. A rather unpleasant proliferation of
Theories Of Everything (TOE). Things are getting worse,
if we just recall the fact that the above five
consistent string theories live in D=10 dimensions, and it is through
compactification that we eventual reach D=4 dimensions. Since the
compactification procedure, even if it is severely constrained, is
rather arbitrary, at least in  perturbation theory, we are landing in
a D=4 landscape made of myriads and myriads of consistent string
vacua. Non-perturbative string effects may help (and do help!), but one
wants to make sure that they don't undo hard-earn successes of
perturbation theory, {\it e.g.} Yukawa couplings, no-scale structures
and the likes.
\item while supersymmetry is a highly desirable symmetry, we better
make sure that it gets broken. Weakly coupled string theory has not
produced a clear-case SUSY breaking mechanism. Even if certain
scenaria, {\it e.g.} gaugino condensation, has been the usual
playground for string phenomenologists, clearly not pinning down the
SUSY breaking mechanism implies not pinning down the SUSY particle
mass spectrum, in other words no {\it hard} string predictions
\item it suffers from an
embarrassing disparity between the observed, apparent scale of (grand)
gauge unification $M_{\rm LEP}$ and the dynamically calculable string
unification scale\cite{Kaplunovsky:1988rp}.
\begin{equation} 
M_{\rm string}\simeq 5 \times g_{\rm GUT} 10^{17}{\rm GeV},
\label{eq:mstring}
\end{equation} 
{\it i.e.} a discrepancy of about a factor of 20.  
Certain ways out in the present framework
(that of weakly coupled strings) have been proposed, including large
threshold effects, extra matter multiplets, non-minimal Kac-Moody
levels and the likes\cite{Dienes:1997du}.  
They all suffer from a common inadequacy.  They
turn a wonderful prediction of the supersymmetric standard model 
(\ref{eq:mgut}) to a
mere fitting of parameters.  On the other hand, we may have yet
another case where non-perturbative string effects may play a
major role.  But how? 
\end{itemize}

Since the, observed at LEP,
superunification of the gauge coupling constants is of such
obvious fundamental importance, it is worthwhile unearthing the
origin of the discrepancy between the $M_{\rm LEP}$ and 
$M_{\rm string}$. The heterotic 
string, the most relevant for phenomenology, will be used as a working
example.  It is well known that\cite{Dienes:1997du}, 
in the heterotic string both gravitational
and gauge interactions are produced from the closed string sector,
thus establishing a relation 
\begin{equation}
G_N\sim \frac{\alpha^{4/3}_{\rm GUT}}{M^2_{\rm GUT}\alpha^{1/3}_{10}}
\label{eq:gn}
\end{equation}
with $\alpha_{10}$ the 10-D string coupling constant, 
while $\alpha_{\rm GUT}$($\equiv\alpha_4$) is the 4-D one.  
It is pretty clear that, with the values of $M_{\rm GUT}$ 
and $\alpha_{\rm GUT}$ as given by (\ref{eq:mgut}) and 
(\ref{eq:alphagut})  respectively, and assuming weakly coupled strings in
10-dimensions ($\alpha_{10} < 1$), we overshoot in the 
``prediction'' of Newton's constant ($G_N$) by about three 
orders of magnitude!  
On the other hand,
if one ``fits in'' the observed value of $G_N$, one gets a value for 
$\alpha_{10}$,
much bigger than 1 indicating that we are really probing the strong
coupling limit of the string in 10-dimensions. 
By recalling the fact that 
\begin{equation}
\alpha_{\rm GUT}\sim \frac{(\alpha')^3\alpha_{10}}{V_6} 
\label{eq:alphagut2}
\end{equation}
with $\alpha'$, the inverse of the string
tension and $V_6$ the compactified 6-D volume, one may entertain the hope
that there is a suitable strong coupling limit, such that 
$\alpha_{\rm GUT} \ll 1$ while
$V_6$, and $\alpha_{10}$ go both to infinity in a suitable manner.  
Then, one may
recover {\it both} the standard prediction of the SSM 
about $M_{\rm GUT}$ and 
the right value of $G_N$. 
In such a case, the strong coupling behavior can be
deduced from what happens in the {\it strongly coupled 10-dimensional
theory}. It is highly remarkable that the bottom-up approach that we
have followed until now has lead us to deduce that the ``vacuum of the
world'' seems to be a ``strongly coupled heterotic string vacuum in
10-D,'' which  has also been the focus of stunning theoretical
developments (top-down approach) the last few years\cite{Polchinski:1996nb}.
So let us shift our attentions to 

\section{Superstrings: Strong Coupling (10-D)}
The strong coupling limit of quantum systems is usually fairly
complicated. Sometimes, we may be lucky and get into the following
situation.  Consider a quantum system $A$ with its fundamental 
degrees of freedom (d.o.f) denoted 
as $X_A$, and let some relevant parameter,
say $g_A$ to go to infinity (strongly coupled limit).  It may happen that
in this limit ($g_A\rightarrow \infty$) 
some of the fundamental d.o.f, $X_A$ turn into
some new d.o.f say $Y_B$, that describe the fundamental d.o.f of another
quantum system $B$, but such that the corresponding relevant parameter
say $g_B$ is much smaller than one (weakly coupled limit).  In such a
situation, by using the mapping $A \leftrightarrow B$, 
we can extract information
about the physics of the strongly coupled $A$ system, by working in the
familiar perturbation regime of relevance to us, of the system $B$!
While all the above may sound and look as a pipe-dream, that is exactly
what is happening in string theory.  Intense theoretical work of the
last few years\cite{Polchinski:1996nb} have indicated that all 
{\it five string theories},
discussed in the previous section, are interrelated in a similar way that 
systems $A$ and $B$ are related above, 
with $g_{A,B}$ referring to the corresponding string
coupling constants.  The magic property that is responsible for all
these correlations is called {\it string duality}.  
This is nothing else but
a non-trivial generalization of the electromagnetic duality, observed
by Dirac, and which lead him to his famous charge quantization
condition\cite{Dirac:1931kp}, in the presence of magnetic monopoles 
\begin{equation}
q_E\cdot q_M = 2\pi n,\quad n= 1, 2,\ldots,
\label{eq:dirac}
\end{equation}
where $q_E$ and $q_M$ refer to electric and magnetic
charges respectively.  In the modern counterpart of non-Abelian gauge
theories, the existence of 
't Hooft-Polyakov type magnetic monopoles, which are not
point-like, but extended objects of solitonic nature resurrected
Dirac's ideas.  Properly modified by Montonen and 
Olive\cite{Montonen:1977sn}, put to work
in supersymmetric Yang-Mills theories.  It was shown 
that in the E-weak limit ($q_E \ll 1$), the solitonic monopoles become
superheavy ($M_{\rm monopole}\propto \frac{1}{q^2_E}$) and thus decouple, 
while in the M-weak limit
($q_M \ll 1$), corresponding through (\ref{eq:dirac}) to $q_E\gg 1$, 
the solitonic
monopoles become massless and provide the new fundamental degree of
freedom!  Actually, extended supergravities, as mentioned in the
introduction, contain naturally non-compact continuous global
symmetries, that act as duality symmetries, that is, exchange
ordinary particles with solitons corresponding to 
{\it weak-strong coupling interchange}. 
Since, string theory yields naturally extended supergravities
in its long-wavelength limit 
({\it e.g.} $N=1$ in $D=10$ $\rightarrow$ $N=4$ in $D=4$), it
shouldn't be that surprising that {\it string duality} 
is present with all
its spectacular consequences.  While, grosso mondo, the generic
analysis above for super Yang-Mills theories hold true, even more
exciting tricks are involved in string theory.  String duality
multiplets may contain, vibrating strings (the basic quanta of
string theory), smooth classical objects of solitonic type, singular
classical objects (black holes) and D-branes\cite{Polchinski:1995mt},
stringy type of
topological defects.  It all depends on the specific string theory and
strong coupling limit chosen, which of the above objects will become
massless and will provide the ``new'' fundamental degree of freedom.  In
10-dimensions, the strong coupling limit of type I ($SO(32)$) string
theory is given by the weakly coupled heterotic $SO(32)$ theory, while
the IIB string is self dual.  Things become more interesting in the
case of IIA strings.  IIA strings in 10-D contain D-0 branes,
topological defects with point-like particle behavior, of mass 
$M_{D-0}\sim \frac{M_{\rm string}}{g}$, 
which in the weakly coupled limit ($g\ll 1$) are superheavy
and leave the vibrating strings to provide the fundamental
d.o.f. Witten\cite{Witten:1995ex} has shown that in IIA strings 
we get towers of $n$ D-0
brane supersymmetric bound states of mass
\begin{equation}
M_{\rm n-th\, bound\, state} \sim n\cdot \frac{M_{str}}{g},
\quad n = 2, 3,\ldots .
\label{eq:boundstate}
\end{equation}
In the strong coupling limit ($g \gg 1$) the D-0 bound states form 
a continuum, recastable as
\begin{equation}
M_n\sim \frac{n}{R_{11}}, \quad R_{11}\equiv\frac{g}{M_{\rm str}}
\label{eq:recast}
\end{equation}
which, of course, is nothing else but the definition of an extra 
{\it 11th-dimension}, \'{a} la Kaluza-Klein!  The
strongly coupled limit of IIA strings is the weakly coupled limit of
some theory in 11-dimensions, compactified on a circle of radius 
$R_{11}$, as given in (\ref{eq:recast}).  This new, 11-D theory is 
called {\it M-theory}\cite{Schwarz:1996bh}.
Actually as Horava and Witten\cite{Horava:1996ma} 
have shown, a different compactification
of M-theory, this time on a semicircle (or segment) 
$\frac{S_1}{Z_2}$ of radius $\rho$,
it provides the strong coupling limit of 10-D heterotic
$E_8\times E'_8$ string, to be discussed in detail later.  
Thus, we see that through
string duality, we have been able to inter-connect all five string
theories in 10-D, and in addition, we lead to the discovery of a new
{\it M(ysterious) Theory} in 11-D.  Of course, as (\ref{eq:recast}) 
indicates, if we
insist on the weakly coupled limit ($g \ll 1$) $R_{11}$, (or $\rho$) 
are much smaller than the string scale ($\frac{1}{M_{str}}$) and 
thus {\it invisible}. Hopefully, all
the above analysis will help understand the string duality motto:
\begin{itemize}
\item The strongly coupled limit of any string theory is 
the weakly coupled limit of another ``string'' theory,
\end{itemize}
with the understanding that ``string'' theory contains also M-theory. 

Non-perturbative string effects, 
attainable through string dualities, have lead to a
much more satisfactory picture of string theory.  Nowadays, all five
string theories in 10-D and the M(ysterious) theory in 11-D are
considered as the limits of {\it one theory}.  In other words, over most of
the theory space, $g \gg 1$, except various specific limits
where $g \ll 1$, corresponding to the above mentioned string theories, and
M-theory.  In fact, currently the 
expression M-theory is used to describe an
unknown, fundamental, 11-D theory, that {\it cannot} formulated as a
traditional quantum theory, due to the lack of a parameter to be
utilized in some perturbation expression, and approximates to 11-D
supergravity at long wavelengths.  A further speculative theory may
exist in {\t twelve-dimensions}, dubbed as {\it F-theory}\cite{Vafa:1996xn},
which gives upon reduction on a two torus the type IIB theory. 
Whatever is the name of the {\it new theory},
it is clear that it resolves one
of the problems of weakly coupled strings, that of the existence of
five different theories in 10-D, by considering them as different
limits of a {\it single} 11-D or 12-D theory.  
On the other hand, the excursion in
11 or 12-dimensions shouldn't be taken lightheartedly, since D=10 has
been advertized as the critical dimension of consistent
superstring theories.  What's going on?  How these extra dimensions
(11 or 12) popped out in string theory? 

While the jury is still out
in answering the above kind of fundamental questions, let me describe
a resolution proposed recently by Ellis, Mavromatos and 
myself\cite{Ellis:1997iw}
in the
framework of the so-called {\it non-critical} or {\it Liouville 
strings}\cite{Ellis:1994vq,Nanopoulos:1994sh}. 
It is well known that string theory contains a degree of freedom, that of
the Liouville filed $\phi$, that {\it decouples} both at the classical and
quantum levels in the critical D=10 superstrings.  It does not
decouple, though, at the quantum level, if we go away from D=10.
Actually, one may employ this new available d.o.f to construct exact
string solutions involving {\it curved spacetimes}, such as a
Robertson-Walker expanding 
Universe\cite{Antoniadis:1988aa,Antoniadis:1989vi,Antoniadis:1991uu},
2-dimensional black-holes\cite{Witten:1991yr}, or to
get new solitonic solutions like NS five-branes\cite{Callan:1991dj} 
and the likes, that
play an important role in non-perturbative string theory\cite{Duff:1995an}.
Sometimes
these solutions are referred to as the {\it linear dilaton} solutions, 
because
the dilaton is proportional to time (cosmological solution) or to some
specific space direction (soliton solution).  Taking into account the
Liouville filed $\phi$ in the world-sheet dynamics, allows for
extrapolation between critical points (corresponding to conformal
field theories), thus providing a way to get out of D=10 and still
obtain a consistent string theory.  We have shown\cite{Ellis:1997iw} 
that in the
Liouville string theory, one may introduce worldsheet topological
defects of the Liouville field, like vortices and monopoles, described
by a deformed sine-Gordon model.  D-branes in target space can be
described in terms of these worldsheet topological defects, and their
connection to black holes become apparent in our 
framework\cite{Ellis:1994vq,Nanopoulos:1994sh}.
The statistical system of vortices and monopoles suffers a
Kosterlitz-Thouless phase transition at a {\it dynamically determined}
critical dimension which turns out to be $D= 11$! Away from
criticality, $D' = D + 1 = 12$, since the Liouville field $\phi$, is not
decoupled.  It is highly tempting to 
identify\cite{Ellis:1997iw}
such an emerging $D=11$
target space theory with M-theory while the corresponding $D' = 12$ with
F-theory!  What such an identification buys us?  Well, to start with,
even if we cannot provide a traditional quantum field
theory in 11 or 12 dimensions, we may have at least a world-sheet
renormalizable theory that represents them.  As we have stressed, 
{\it in dissent} and {\it for several 
years}\cite{Ellis:1994vq,Nanopoulos:1994sh},
we may be in for surprises on the form
of quantum field theory that is descenting from string theory.  A
further indication of the real worth of the Liouville string
framework, is provided by contemplating on the well known fact that
the maximum number of dimensions in which supersymmetry can exist is
$D' = 12$, provided that the signature of spacetime is (10,2).  The tale
of two times!  Our proposal\cite{Ellis:1997iw} 
for managing the appearance of these two
times is to identify the zero modes of the {\it a priori} distinct quantum
fields $X^0$($\equiv t$ in critical strings) and $\phi$, 
in the neighborhood of each fixed point corresponding to the five string
theories, and the 11-dimensional theory. In other words, each fixed point
in the space of ``string'' theories has its own time-like
coordinate $X^0$, which is a {\it reversible} coordinate, in the Einstein
sense, so that a fixed point is Lorents invariant.  In addition, in
the bulk of the theory space there is a second time-like coordinate,
the Liouville field $\phi$, with respect to which evolution is
{\it irreversible} in general.  This means that the general 
12-dimensional target space F-theory is ``non-equilibrium'' and does not
have a simple field-theoretic interpretation.  Thus it is possible to
turn an apparent embarrassment of riches, that of signature (10,2) to
our advantage and resolve a hundred year mystery, that of the
(microscopic) origin of the arrow of time, while keeping Einstein
physics basically intact.  Last and not least, our approach
reestablishes the singular importance of strings.  Maybe, {\it 
L(iouville) theory will come to be known as the theory formerly 
known as M/F theory}.

\section{Strongly coupled $E_8\times E'_8 \rightarrow$ M-theory: A
phenomenological profile}
The heterotic $E_8\times E'_8$ string, in its
weakly coupled form has been the focus of intense phenomelogical
studies, due to its rich structure that yields realistic models
describing the world at long wavelengths.  In the strong coupling
limit, $E_8\times E'_8$ emerges with certain unique characteristics,
that once more make it a very promising framework to study
phenomenology. One may, up front, question the meaning of the whole
exercise, by noticing that we don't know what M-theory is, so what we
are talking about, its compactification and the likes?  The idea here is
the following: we do know that the infrared limit of M-theory has to
be {\it $11-D$ superqravity}, 
since there is no other consistent quantum field
theory available!  Furthermore, since $11-D$ supergravity (or $5-D$
supergravity) are fairly well-studied, we may get quite a lot of
information about M-theory, by studying $11-D$ SUGRA, if it happens 
that $\rho$, the 11-th (or 5-th) compactified dimension is 
much bigger than the 11-th Planck scale ($\frac{1}{M_{11}}$). 
As we are going to see soon,
phenomenological requirements put us in a 
$\rho\cdot M_{11} \gg {\cal O}(1)$ regime and thus enabling us 
to study {\it M-phenomenology}.
The low-energy consequences of the unknown, at microscopic level 
($\sim M^{-1}_{11}$), M-theory may be unveiled by studying suitably 
tailored $11-D$ SUGRA (or $5-D$ SUGRA).  After all, Fermi's theory of 
$\beta$-decay, augmented
with parity violation and Cabibbo currents, provided a pretty good
picture of low energy weak interactions ($\gg M^{-1}_W$), 
we didn't have to
wait till the '70s, when a microscopic ($\sim M^{-1}_W$) 
theory of electroweak interaction was available, 
in order to study low energy weak
interactions!  

As discussed in the previous section, the strongly
coupled $10-D$ $E_8\times E'_8$, is described by M-theory compactified on a
segment (or semicircle) $\frac{S_1}{Z_2}$, 
of dimension $\rho$\cite{Horava:1996ma}. Of course,
realistically one considers $R_4 \otimes X_{CY}$ compactifications 
of the strongly
coupled $E_8\times E'_8$, which corresponds to a 
$R_4 \otimes X_{CY}\otimes \frac{S_1}{Z_2}$
compactification of M-theory, with $X_{CY}$ denoting an
appropriate Calabi-Yau threefold (= 6 space dimensions).  The
11-th dimension ($\rho$) has an orbifold structure 
$\frac{S_1}{Z_2}$ that is instrumental: 
\begin{itemize}
\item at one end live the observable fields contained
in $E_8$, at the other end live the hidden sector fields contained in
$E'_8$, and in the middle (``bulk'') propagate the gravitational fields.
The fields that live on the two boundaries, may be considered as
comprising the ``twisted'' sector, while the gravitational
fields in the bulk make up the ``untwisted'' sector. 
\item as such,
the fields living at the boundaries, are {\it oblivious} to the existence of
the 11-th dimension ($\rho$), whatever is its compactification scale!  Such
a property has far-reaching phenomenological consequences:
\begin{enumerate}
\item Since the observable 
fields live in the $10-D$ ($4-D$ after compactification)
boundaries, {\it chirality} is not an issue, in sharp contrast to
conventional manifold compactifications of $11-D$ SUGRA where it is
fatal.
\item There are no Kaluza-Klein towers of particles, based on
the boundary living fields.  Thus, a standard severe problem of the
``large compactification radius'' type models\cite{Antoniadis:1990ew}, 
that of the breakdown of
perturbation theory at the compactification radius ($\gg M^{-1}_{\rm GUT}$),
is naturally evaded.  Gauge unification, and for that matter, Yukawa
coupling unification proceeds normally as in the SSM, {\it i.e.}, 
$M_{\rm GUT} = M_{\rm LEP}$
independent of the size of $\rho$!
\end{enumerate}
\end{itemize}
Furthermore, one may naturally
identify the 11-th dimensional Planck mass ($M_{11}$) with 
$M_{\rm LEP}$, which also
provides the characteristic Calabi-Yau compactification scale.  One
then finds that (\ref{eq:gn}) is replaced by\cite{Witten:1996mz,Li:1997ii}
\begin{equation}
G_N = \frac{1}{4}\left(\frac{\alpha_{\rm GUT}}{M^2_{\rm LEP}}\right)
\left(\frac{\rho^{-1}_0}{M_{\rm LEP}}\right)
\label{eq:gn2}
\end{equation}
which for $\rho_0$, the compactification radius of the
11-th dimension,
\begin{equation}
\rho^{-1}_0\sim (10^{12}-10^{13}){\rm GeV}
\label{eq:rho0}
\end{equation}
gives the right value for $G_N$!  
Of course what we have realized here\cite{Witten:1996mz} is what we
have foresaw in section 2, but with a ``twist''.  Indeed, the existence
of an extra 11-th dimension $\rho$, has allowed us to enlarge suitably the
compactified volume ($V_6\times\frac{S_1}{Z_2}$), 
such that $\alpha_{\rm GUT}$
remains much smaller than one, while at the same time the 11-th
dimension is large enough (see (\ref{eq:rho0})) 
so that an 11-D SUGRA approach is
justifiable!  At an intuitive level what is really happening is the
following.  While the three gauge coupling constants evolve
dynamically with energy, and meet at $M_{\rm LEP}$, because they are 
{\it 11-th dimension blind}, 
the gravitational constant after we hit $\rho_0$, it is
replaced by a 5-dimensional one, so that for $E > \rho^{-1}_0$, instead of
considering $G_N E^2$ as the dimensionless relevant constant, we
ought to work with $G_N E^2(\frac{E}{\rho^{-1}_0})$ that increases 
much faster with energy and
enables unification of all interactions at $M_{\rm LEP}$!  
The reader may
have already noticed that all the above marvelous picture depends on
the specific value of $\rho^{-1}_0$ as given by (\ref{eq:rho0}).  
In order to claim that
we have resolved the GUT scale-string scale disparity problem
(see section 2), we need to determine {\it dynamically} the value of 
$\rho^{-1}_0$, and hopefully it will be given still by (\ref{eq:rho0}).  
This issue brings us naturally to our next point.

The scalar potential $V$ is {\it independent}
of $\rho$, thus it contains a {\it flat direction}, 
at the classical level, that
may serve as the basis for implementing the {\it no-scale supergravity
framework} (\ref{eq:m32})\cite{Lahanas:1987uc}.  
Indeed, several groups\cite{Li:1997ii,Banks:1996ss,Horava:1996vs,Li:1997sk,Antoniadis:1996hk,Dudas:1997jn}
have reached the same
conclusion, namely that no-scale supergravity is the long wavelength
limit of M-theory in 4-dimensions.  In particular, we provided the
first explicit calculation\cite{Li:1997sk} 
that supports the above remarks.  Upon
compactification of M-theory, in its 11-D SUGRA form, on a Calabi-Yau
manifold with Hodge numbers $h_{(1, 1)} = 1$ and $h_{(2, 1)} = 0$ 
and boundary $\frac{S_1}{Z_2}$, 
a no-scale K\"ahler potential, superpotential and gauge
kinetic function were obtained explicitly\cite{Li:1997sk}.
In four dimensions, this
result is related to the previous weakly-coupled string no-scale
supergravity result, obtained by Witten\cite{Witten:1985xb}, through a
field transformation, which means that they are equivalent in
4-dimensions.  This robust behavior of the no-scale supergravity
framework\cite{Lahanas:1987uc}, 
all the way from the weakly coupled to strongly coupled
heterotic string is rather remarkable, and highly suggestive that 
(\ref{eq:m32})
may be implemented at the phenomenologically relevant strong coupling
limit of $E_8\times E'_8$ heterotic string. After all, no-scale 
supergravity traces its 
origins\cite{Cremmer:1983bf,Ellis:1984ei,Ellis:1984bm} in the non-compact 
continous global symmetries of extended supergravities, whose subgroups, suitably treated, provide {\it string duality}!

Supersymmetry breaking
may be supplemented at the string level by 
employing\cite{Li:1997ii,Li:1997sk,Antoniadis:1996hk,Dudas:1997jn}
the Scherk-Schwarz (SS) mechanism\cite{Scherk:1979zr}
on the 11-th dimension.  As is well known, the Scherk-Schwarz
SUSY breaking
mechanism makes use of a symmetry of the theory transforming the
gravitino nontrivially.  In our case, since the 5-th dimension is
compactified on $\frac{S_1}{Z_2}$, 
the symmetry under discussion must be a $2\pi$
rotation on the plane of the 5-th dimension and one of the
internal Calabi-Yau coordinates.  In a way, such a symmetry acts on
the 5-D fields as the space-time parity $(-1)^{2s}$, {\it i.e.}, 
changes sign for
the fermions and leave bosons invariant\cite{Antoniadis:1996hk}. Thus
\begin{equation}
m_{3/2}=\frac{1}{2\rho}.
\label{eq:m32new}
\end{equation}
Notice that, it is only the fermions in the ``bulk'' (``untwisted''
sector) that receive a uniform shift in their $p_5$ momentum 
($\sim (n+1/2)\rho^{-1}$, while  both fermions and bosons on 
the boundaries (``twisted'' sector), 
as living on the semicircle edges have no $p_5$ momenta and thus
no supersymmetry breaking contributions.  Supersymmetry breaking will,
then, be communicated from the ``bulk'' by gravitational interactions.  It
is highly amazing, how closely the above scenario resembles the 
{\it no-scale framework}\cite{Lahanas:1987uc}.
Indeed, the SS mechanism provides a flat potential
(no-cosmological constant at the classical level) along the 11-th (or
5-th) dimension $\rho$, with SUSY breaking at the classical level (see
(\ref{eq:m32new})), but with the magnitude of the SUSY breaking 
(alias gravitino mass)
{\it undetermined} at the classical level.  
But this is the no-scale SUGRA
framework!  We then can employ quantum corrections to {\it dynamically}
determine everything, \'{a} la (\ref{eq:m32}), including the magnitude of the
compactification scale of the 11-th (or 5-th) dimension $\rho_0$, as
promised above.  In fact, one may see that what we expect to get out
is in the right ballpark, {\it i.e.} (\ref{eq:rho0}) 
gets satisfied.  Indeed, one expects
naively that the  no-scale mechanism will dynamically fix, as usual, the
amount of the observable SUSY breaking scale, relevant for the gauge
hierarchy problem, at $\tilde{m}\sim {\cal O}(1 {\rm TeV})$,
which  is related to $m_{3/2}$ by $\frac{m^2_{3/2}}{M}$,
with $M$ some scale in the ($10^{16} - 10^{18} {\rm GeV}$) range, 
and thus fixing dynamically $m_{3/2}$, and thus $\rho^{-1}$ through 
(\ref{eq:m32new}) to be in the
right domain, given by (\ref{eq:rho0})!  
Work in progress\cite{Benakli:1997aa} indicates that such an
optimistic scenario is not far from reality.  All the above results
depend critically on the {\it stability} of the no-scale framework, which in
turn depends strongly on the fact, that there are no quadratic
divergences in the effective supergravity.  Indeed, another, of utmost
importance, result of the specific SS  SUSY breaking mechanism considered
here is the vanishing of Str$M^2$ after supersymmetry 
breaking\cite{Antoniadis:1996hk}. 

The communication of supersymmetry breaking to the observable sector,
attached to one of the boundaries, from the ``bulk'', where it was
originated, through gravitational interactions, is rather intrigued and
maybe not very clear, at the time of writing.  Horava\cite{Horava:1996vs} 
has argued that
supersymmetry breaking ($m_{3/2}\ne 0$) is not felt immediately in the
observable sector because of a topological obstruction (essentially
the 11-th dimension of length $\rho_0$ that separates the two sectors).  In
fact he has argued\cite{Horava:1996vs} 
that there is a hidden 11-D supersymmetry, broken
only by the {\it global topology} of the {\it orbifold dimension} 
($\rho_0$), that
explains the ``conspiracy'' that leads in the weakly coupled heterotic
string theory to the {\it no-scale structure}, with SUSY breaking and
vanishing cosmological constant, at the classical level.
Supersymmetry breaking becomes apparent only after the renormalization
scale is low enough to not reveal the presence of the 11-th dimension
anymore.  In practice, one is to allow for non-vanishing SUSY breaking
parameters only for scales $Q < \rho^{-1}_0$\cite{Li:1997ii}.
A similar conclusion can be reached 
by looking in the dual, strongly coupled 
$E_8\times E'_8$ theory, where as
Witten has shown\cite{Witten:1996mz},
one reaches a strongly coupled $E'_8 (a_{8'} > 1)$, suitable
for gaugino condensation, only when $\rho$ gets some critical
value, $\rho_{\rm crit}$.  
In fact, $\rho_{\rm crit}\sim\frac{\alpha_{\rm GUT}}{16\pi^2}M_{\rm LEP}$, 
very close, at
least numerically, to $\rho_0^{-1}$, as given by (\ref{eq:rho0}). 
Thus, once more $Q <\rho_{\rm crit}^{-1}$
in order to ``feel''  the SUSY breaking in the observable
sector.  This effect can leave a deep imprint on the low energy
sparticle spectrum, which depends quantitatively on the amount of
``running'' of these parameters\cite{Li:1997ii}. 
Until now I have tried to
present general characteristics of the anticipated M-phenomenology,
without resorting to a specific scenario or model. In order to get
some experimentally testable predictions, we need now to be more
specific in the selection of boundary conditions for the softly broken
parameters at $\Lambda_{\rm SUSY}\sim\rho^{-1}_0$\cite{Li:1997ii}.  
Opinions are divided on this issue, and it
is fair to say that things are not yet crystal clear.  We have
chosen\cite{Li:1997ii} to take $m_0 = 0$, at $\Lambda_{\rm SUSY}$, 
as it avoids FCNC problems and make
the effect of taking $\Lambda_{\rm SUSY}\sim\rho^{-1}_0$ 
most noticeable.  While this choice
arises in certain scenaria, cannot be claimed to be indispensible.
Anyways, let us see what type of SUSY phenomenology is coming out.

\section{M-theory Inspired Supersymmetry phenomenology}
\begin{figure}[p]
\includegraphics{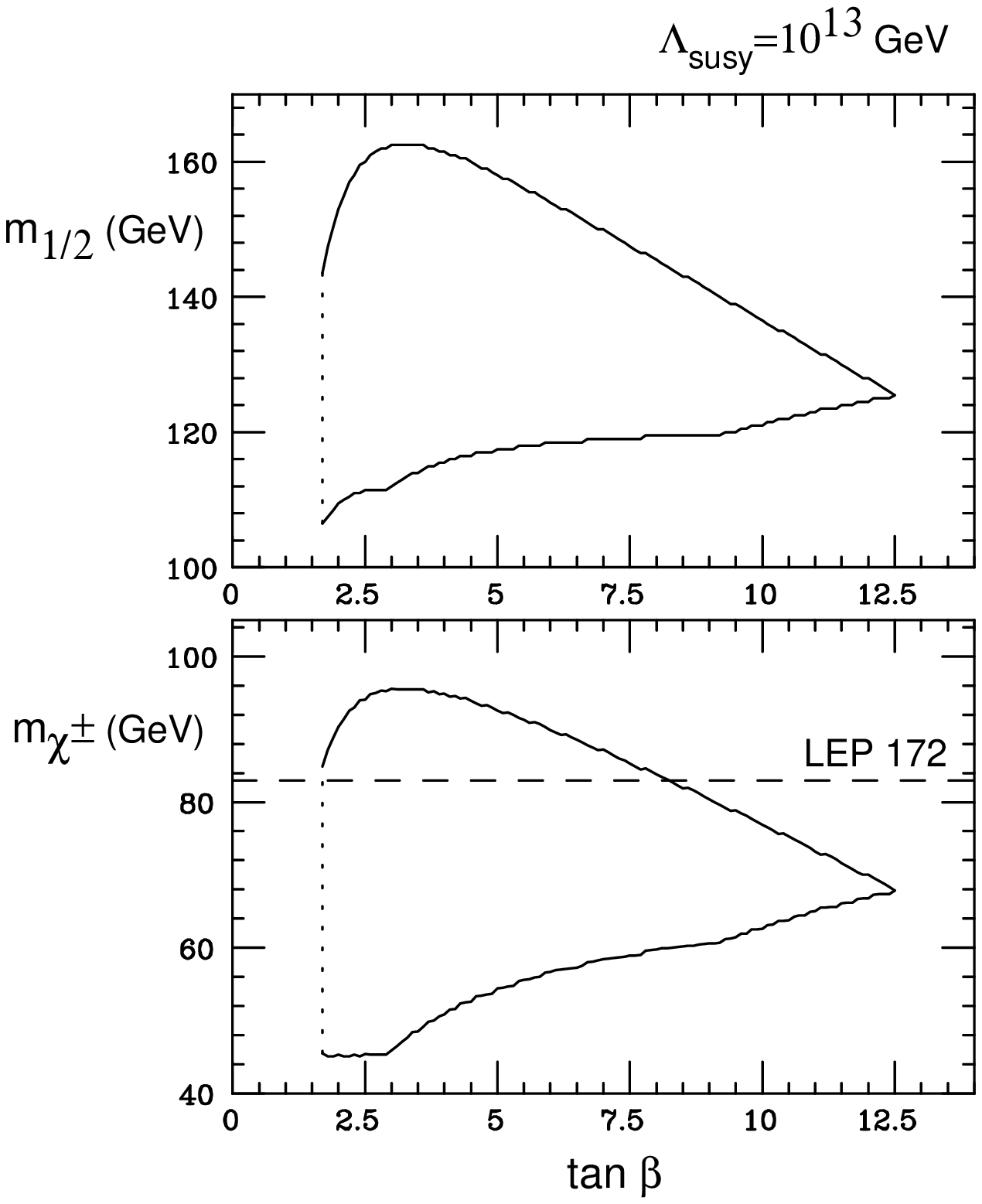}
\vspace{5.5in}
\caption{The allowed region in $(m_{1/2},\tan\beta)$ [top panel] and
correspondingly $(m_{\chi^\pm},\tan\beta)$ [bottom panel] in no-scale
supergravity ($m_0=A_0=0$) with $\Lambda_{\rm susy}=10^{13}\,{\rm GeV}$.
Above the top boundary $m_{\tilde e_R}\approx m_{\tilde\tau_1}<m_\chi$, 
whereas below the bottom boundary $m_{\tilde\nu}<43\,{\rm GeV}$. 
The dashed line [bottom panel] represents the lower bound on the 
chargino mass from LEP~172 searches.}
\label{fig:boundary13}
\end{figure}

We now proceed to the analysis of the low-energy sparticle 
spectrum under the
assumptions of $\Lambda_{\rm susy}=\rho^{-1}_0$ and $m_0=A_0=0$.
Thus, the only free parameters are $m_{1/2}$ and $\tan\beta$. We find that
the requirement of radiative electroweak symmetry breaking plus two basic
phenomenological requirements, allow solutions in the $(m_{1/2},\tan\beta)$
plane for only one sign of $\mu$ and only within a completely bounded region.
For the case of $\Lambda_{\rm susy}=10^{13}\,{\rm GeV}$, this region is
shown in Fig.~\ref{fig:boundary13}, where to facilitate comparison with
experiment we also show the region in the $(m_{\chi^\pm},\tan\beta)$ plane. 
The upper limit on $m_{1/2}$ (for a fixed value of $\tan\beta$) follows from
the requirement that the lightest supersymmetric particle be 
neutral\cite{Kelley:1993bu}.
Above the upper boundary the right-handed selectron ($\tilde e_R$) 
becomes lighter than the lightest neutralino ($\chi$).
The bottom boundary is obtained by imposing the absolute lower 
limit on the sneutrino mass from LEP~1 searches 
($m_{\tilde\nu}>43\,{\rm GeV}$). 
The area to the right of the right-most tip of the region is 
excluded by these two conflicting constraints. 
The $\tan\beta$ dependence of these constraints may be 
understood from the D-term contribution to the $\tilde e_R$
and $\tilde\nu$ mass formulas
\begin{equation}
\widetilde m^2_i=c_i\, m^2_{1/2} - d_i\, 
{\tan^2\beta-1\over\tan^2\beta+1}\, M^2_W\ ,
\label{eq:masses}
\end{equation}
where the $c_i$ are some RGE-dependent constants and 
$d_{\tilde e_R}=-\tan^2\theta_W<0$ whereas 
$d_{\tilde\nu}={1\over2}(1+\tan^2\theta_W)>0$.
The dotted line indicates the lower bound on $\tan\beta$ 
that is consistent with the top-quark mass ($m_t=175\,{\rm GeV}$) 
and perturbative Yukawa couplings up to the unification 
scale. In practice, the LEP~172 lower 
bound on the chargino mass ($m_{\chi^\pm}>83\,{\rm GeV}$)\cite{ALEPH}
gives the strongest constraint on the parameter space 
(dashed line on bottom panel in Fig.~\ref{fig:boundary13}). 
Nonetheless, a portion of the parameter space remains allowed, 
and in fact it is within the reach of future LEP~2 energy upgrades, 
as we discuss below.

\begin{figure}[p]
\includegraphics{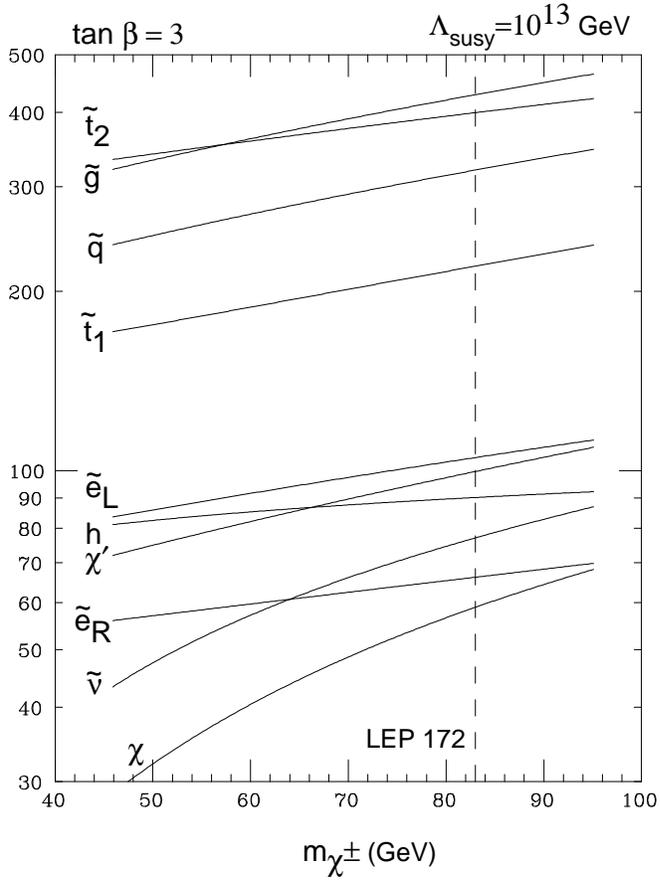}
\vspace{5.5in}
\caption{Calculated values of representative sparticle masses 
versus the chargino mass for 
$\Lambda_{\rm susy}=10^{13}\,{\rm GeV}$ and $\tan\beta=3$.
The spectrum terminates when $m_\chi$ approaches $m_{\tilde e_R}$ from below.
The dashed line represents the lower bound on the chargino mass 
from LEP~172 searches.}
\label{fig:spectrum}
\end{figure}

To give a more detailed picture of the low-energy spectrum, 
in Fig.~\ref{fig:spectrum} we display representative sparticle masses 
as a function of the chargino mass for 
$\Lambda_{\rm SUSY}=10^{13}\,{\rm GeV}$ and
$\tan\beta=3$. 
This choice of $\tan\beta$ allows the widest range of sparticle
masses (see Fig.~\ref{fig:boundary13}). This figure shows that the spectrum
``terminates'' when $m_\chi$ approaches 
$m_{\tilde e_R}$ from below, as mentioned
above in connection with the upper boundary in Fig.~\ref{fig:boundary13}. 
It is
interesting to note the significant splitting of 
the top-squark ($\tilde t_{1,2}$) masses around the 
average squark ($\tilde q$) mass.

In the LEP 172 allowed region in Figs.~\ref{fig:boundary13} and 
\ref{fig:spectrum} we find $m_{\chi^\pm_1}<95\,{\rm GeV}$ and 
$m_{\tilde e_R}<70\,{\rm GeV}$. 
Both of these particles appear within 
the reach of LEP~2. More to the point, one might
wonder whether the such light right-handed selectron masses might have already
been excluded by LEP~2 searches, 
as they have been certainly kinematically accessible. 
We have calculated the cross section 
$\sigma(e^+e^-\to\tilde e^+_R\tilde e^-_R)$ 
at LEP~161, for which explicit limits have been released by
the OPAL Collaboration\cite{Ackerstaff:1997aw}.
We find $\sigma<0.2\,{\rm pb}$, which in
${\cal L}=10.1\,{\rm pb}^{-1}$ would have yielded a maximum of two events.
Indeed, the experimental sensitivity to this mode is at the 0.5~pb 
level\cite{Ackerstaff:1997aw}.
Thus, past LEP~2 searches in 
the selectron channels do not restrict the allowed 
parameter space any further. Moreover, near the
upper end of the parameter space the experimental 
detection efficiency should
be greatly reduced because $m_{\tilde e_R}$ approaches $m_\chi$. 
One should also consider the predictions for trilepton events at the 
Tevatron. We find 
$\sigma(p\bar p\to\chi^\pm\chi')\approx (1.0-0.7)\,{\rm pb}$ for 
$m_{\chi^\pm}=(83-95)\,{\rm GeV}$. 
The leptonic decays of the chargino and
neutralino are maximally enhanced because of the lighter 
right-handed sleptons
and sneutrinos, respectively. That is, 
$B(\chi^\pm\to\ell\nu_\ell\chi)\approx 2/3$
and $B(\chi'\to\ell^+\ell^-\chi)\approx1/2$, 
where $\ell=e+\mu$. Combining these numbers we arrive at a single 
channel ({\em i.e.}, any single one of
$eee$, $ee\mu$, $e\mu\mu$, or $\mu\mu\mu$) cross section of $(0.16-0.11)$~pb.
This result is slightly below the sensitivity reached at the Tevatron in
trilepton searches\cite{Abbott:1997ri}, 
and thus these also do not constrain the
allowed parameter space any further.

\section{Conclusions}
Strongly coupled strings, as
studied by the use of string duality, seem to provide one {\it single
theory} and hold the potential to lead us to a unique string vacuum,
hopefully involving $E_8\times E'_8$. Stringy new M-theory may cause a
{\it paradigm-shift} in the way we are understanding low-energy physics.  
A new way of understanding gauge-gravitational unification is suggested
and already put to work, a new way of SUSY breaking is emerged, which,
while contains seeds of the past, it resolves several severe problems
and may lead eventually, into a clear-cut SUSY spectrum, among other
things, that may provide a smoking gun for physics at the (11-th
dimension) Planck scale. For example, a deeper understanding of proton
stability ($\tau_p > 10^{33}{\rm years}$) in SUSY theories 
may be needed, and such
constraints may reduce considerably the class of acceptable
M-compactifications\cite{Ellis:1997ec}. 
On a different wavelength, black hole dynamics may
be studied explicitly and quantum mechanics may suffer modifications,
similar to the ones occurred to the Newtonian
gravity after the innocent looking {\it equivalence principle} was
implemented correctly by Einstein.  Here, duality symmetries are
realized {\it only} at the quantum level, and while, once more, innocent
looking, may carry the seeds to a complete revision of quantum 
theory\cite{Witten:1997yq}.
It is not, yet, judgment day, it is only the beginning excellent and
fair$\ldots$

\section*{Acknowledgments}
This work has been supported by DOE grant DE-FG03-95-ER-40917.

\bibliography{dimitri,misc}
\bibliographystyle{prsty}

\newpage

\newpage

\end{document}